\documentstyle[pra,aps,twocolumn]{revtex}

\author{ M. A. Nielsen \thanks{Electronic address : mnielsen@tangelo.phys.unm.edu}}
\title{The entanglement fidelity and quantum error correction}
\address{Center for Advanced Studies, Department of Physics and
Astronomy, University of New Mexico, Albuquerque NM 87131-1156}

\date{\today}

\begin{document}


\maketitle

\begin{abstract}
Two new expressions for the
{\em entanglement fidelity} recently introduced by Schumacher
(LANL e-print quant-ph/9604023, to appear in Phys.\ Rev.\ A) are derived. These
expressions show that it is the entanglement fidelity which must be
maximized when performing error correction on qubits for quantum computers,
not the fidelity, which is the most-often used generalization of the
probability for storing a qubit correctly.
\end{abstract}


There is great interest in the use of entangled quantum states as
a resource to accomplish
tasks which are impossible or difficult to do classically. These tasks
include quantum cryptography \cite{Bennett92a}, quantum
teleportation \cite{Bennett93a},
quantum coding \cite{Schumacher95a,Jozsa94a,Barnum96b}
and quantum computation \cite{DiVincenzo95a,Bennett95a}. For each of
these tasks it is desirable to be able to maintain the entanglement of
a subsystem with the remainder of the system.

A great deal of work has recently been done on quantum error correction;
a sample of this work can be found in
\cite{Shor95a,Ekert96a,Knill96a,Plenio96a,Schumacher96b,Shor96a,Steane96a}.
Much of this work has focused on preserving the
state of a quantum system, without explicitly taking into
account whether entanglement is preserved. Later in this Rapid Communication
an example is given where the quantum state is preserved, while the
entanglement is completely destroyed. First, though, we review what it means
for the quantum state alone to be preserved, without accounting for
entanglement.

Suppose, for example, that a single qubit is to be stored in the
memory of a quantum computer. The qubit starts out in a pure state
$|\psi\rangle$, and sometime later the interaction of the system with its
environment has caused the state of the qubit to change to $\rho$. An obvious
way to quantify how well the state has been stored is the quantity
\begin{eqnarray}
P_e = 1 - \langle \psi | \rho |\psi \rangle, \end{eqnarray}
which measures the probability of error in storage of the qubit. 
It would seem that this is the natural quantity to minimize in order
to perform good quantum error correction. Equivalently, we wish to
maximize the probability the state is stored correctly,
\begin{eqnarray} \label{pr correct}
P_c = \langle \psi |\rho |\psi \rangle. \end{eqnarray}
Clearly $P_c$ is $1$ if and only if the state is stored correctly, and
less than $1$ if some error has occurred during storage.

More generally we would like to have a measure of how successfully a mixed
quantum state is stored.
Suppose the initial quantum state is $\rho_1$ and the
final quantum state is $\rho_2$.
What does it mean to say that these states are different, that is,
can be distinguished? One measure of distinguishability with very useful
properties is known as the {\em fidelity} \cite{Jozsa95a}
between the two states. It can be
defined by the expression
\begin{eqnarray} \label{defn fidelity}
F(\rho_1,\rho_2) := \max |\langle \psi_1| \psi_2 \rangle|^2, \end{eqnarray}
where the maximum is taken over all purifications \cite{Jozsa95a},
$|\psi_1\rangle$ and $|\psi_2\rangle$, of the states $\rho_1$ and $\rho_2$.

One useful property of the fidelity is that it reduces to $P_c$ in the
case when $\rho_1$ is a pure state,
\begin{eqnarray}
F \Bigl ( |\psi_1\rangle\langle \psi_1|, \rho_2 \Bigr )
	 = \langle \psi_1| \rho_2 |\psi_1 \rangle. \end{eqnarray}
The fidelity has many other desirable properties \cite{Jozsa95a}. It
is obviously symmetric in $\rho_1$ and $\rho_2$. It satisfies the
inequality
\begin{eqnarray}
0 \leq F(\rho_1,\rho_2) \leq 1, \end{eqnarray}
and $F(\rho_1,\rho_2) = 1$ if and only if $\rho_1 = \rho_2$. The
smaller the fidelity, the more distinguishable the states
are. These properties, together with the agreement with (\ref{pr correct}) in
the pure state case make the fidelity an attractive measure of how well
a quantum state has been preserved.

The next concept we introduce is that of an {\em extension} $\tilde \rho$
of $\rho$. Suppose $\rho$ is defined on a system $S$ which is part of a
larger system $ST$ which has density operator $\tilde \rho$, that is,
\begin{eqnarray}
\mbox{tr}_T(\tilde \rho) = \rho, \end{eqnarray}
where $\mbox{tr}_T$ denotes the partial trace over $T$. Any density operator
$\tilde \rho$ satisfying this condition will be called an extension of
$\rho$. Notice that a purification $|\psi\rangle$ of $\rho$ gives rise
to an extension $|\psi\rangle\langle \psi|$ of $\rho$. 

It follows easily from the definition of fidelity that if
$\tilde \rho_1$ and $\tilde \rho_2$ are extensions of $\rho_1$ and
$\rho_2$ to the same larger system $ST$, then \cite{Jozsa95a}
\begin{eqnarray} \label{non-decrease trace}
F(\tilde \rho_1,\tilde \rho_2) \leq F(\rho_1,\rho_2). \end{eqnarray}
Physically what this means is that the density operators for the
subsystem $S$ are less distinguishable than those for the whole
system $ST$, which is what we expect, since having access to
the total system should not make it any more difficult to distinguish the
states.

One final useful property of fidelity concerns quantum operations.
A quantum operation is the most general physically reasonable map
that can be used to represent the change in a quantum state during
interaction with an environment. Strictly speaking we are only
concerned here with {\em non-selective} quantum operations in which
the result of any measurement on the environment performed after the
interaction is disregarded, that is, we are only concerned with the
non-selective evolution of the quantum state.
A detailed description of quantum operations can be found in
\cite{Kraus83a} and a more physical account in the appendix to
\cite{Schumacher96a}. For our purposes we will need two (equivalent)
ways of representing quantum operations. The first is the representation
in terms of unitary operations with an ancilla. A quantum operation can
always be represented by introducing an ancilla system $E$, and a unitary
operator $U$ on the system plus ancilla, $SE$, in such a way that the
operation has the form
\begin{eqnarray} \label{unitary rep}
{\cal E}(\rho) = \mbox{tr}_E \bigl ( U (\rho \otimes \sigma) U^{\dagger}
	\bigr ), \end{eqnarray}
where $\mbox{tr}_E$ denotes tracing out the ancilla, and $\sigma$
is the initial state of the ancilla. Conversely, any map of this
form is a quantum operation.

The second representation for a quantum operation is the
{\em operator sum\/} representation. In general any quantum operation
can be written in the form
\begin{eqnarray} \label{operator sum rep}
{\cal E}(\rho) = \sum_i A_i \rho A_i^{\dagger}, \end{eqnarray}
where the $A_i$ are system operators satisfying the completeness relation
$\sum_i A_i^{\dagger} A_i = I$. Conversely, any map of this
form is a quantum operation.

How does the fidelity behave under quantum operations? It is not difficult
to show \cite{Barnum96a} from (\ref{defn fidelity}),
(\ref{non-decrease trace}) and (\ref{unitary rep}) that
\begin{eqnarray} \label{non-decrease evop}
F \Bigl ( {\cal E}(\rho_1),{\cal E}(\rho_2) \Bigr )
	 \geq F(\rho_1,\rho_2). \end{eqnarray}
That is, a quantum operation can only increase the fidelity between
two states. This inequality is intuitively appealing since it says that
physical processes can not increase the distinguishability of
quantum states.

The fidelity is a very useful tool for analyzing the storage of states,
but it does not take into account the possible entanglement of a system
with other systems. A simple example to illustrate this point involves
the storage of two two-level systems by Alice and Bob. In this example
we will be considering Alice's system as the subsystem of interest, analogous
to a single qubit of memory, while
Bob is analogous to the remainder of the quantum computer.

Consider two possible dynamics for the combined system. The
first is perfect storage of the entire system. If the initial density operator
of the combined system is labelled $\rho$ then this dynamics is given by the
map
\begin{eqnarray}
\rho \rightarrow \rho ' = \rho. \end{eqnarray}
Note that this map is clearly a quantum operation.

The second dynamics we will consider destroys $\rho$ and leaves the
system in a completely mixed state,
\begin{eqnarray}
\rho \rightarrow \rho ' = \frac{I \otimes I}{4}. \end{eqnarray}
To see that this is a quantum operation we will use the representation
(\ref{unitary rep}) for quantum operations. Introduce an ancilla system
consisting of two two-level systems (``Ted'' and ``Carol''). This
ancilla is started in the state $\sigma = I \otimes I / 4$. Suppose $U$
is the unitary operator which swaps the state of system $AB$ with
the state of system $TC$. Then
\begin{eqnarray}
\rho \rightarrow \mbox{tr}_{TC}(U (\rho \otimes \sigma) U^{\dagger}) = 
	\frac{I \otimes I}{4} \end{eqnarray}
is a quantum operation.

For Alice's system alone the corresponding dynamics are given by the
quantum operations,
\begin{eqnarray}
\rho_A \rightarrow {\cal E}_1(\rho_A) = \rho_A, \end{eqnarray}
and
\begin{eqnarray}
\rho_A \rightarrow {\cal E}_2(\rho_A) = \frac{I}{2}. \end{eqnarray}

Suppose now that Alice and Bob start with a shared EPR pair,
\begin{eqnarray}
|\psi\rangle = \frac{|\uparrow \downarrow \rangle -
	 |\downarrow \uparrow\rangle}{\sqrt{2}}. \end{eqnarray}
The state of Alice's system alone is initially
\begin{eqnarray}
\rho_A = \mbox{tr}_B(|\psi\rangle \langle \psi|) = \frac{I}{2}, \end{eqnarray}
where $\mbox{tr}_B$ indicates a partial trace over Bob's system.
Notice that under either dynamics the final state of Alice's system
is given by
\begin{eqnarray}
\rho_A ' = \frac{I}{2}. \end{eqnarray}
Thus under either dynamics Alice's system has been stored perfectly,
that is, with fidelity equal to $1$.
However, the first dynamics leaves the entanglement of Alice's system with
Bob's intact, while the second dynamics destroys
the entanglement. Clearly, if we are interested in using entanglement as
a resource, fidelity alone is not a sufficient measure of how well a
quantum system is stored.

One could argue that what should be done is to look at the fidelity of the
combined system belonging to Alice and Bob - this may be feasible in this
simple example. However, in general, quantum computers can be very large
systems compared to the subsystem (analogous to Alice's system) whose
performance as a memory element we wish to analyze, and inclusion of the
entire state and dynamics of the quantum computer would make the analysis
enormously complicated.  What this example shows
is that the fidelity of the subsystem density operators is not the
correct quantity to look at to analyze the
performance of the subsystem as a storage device if storing entanglement is
important.
 
We will now define a quantity analogous to fidelity which does keep track of
how well the state and entanglement of a subsystem of a larger
system are stored, without requiring that the complete state or dynamics of the
larger system be known. We will then prove that this quantity is equal to
the {\em entanglement fidelity} defined by Schumacher \cite{Schumacher96a}.
Let us define a quantity, $F_1$, by the expression
\begin{eqnarray}
F_1(\rho, {\cal E}) := \min_{\tilde \rho, {\cal E}'}
	 F \Bigl ( ({\cal I}_S \otimes {\cal E} ') (\tilde \rho),
	({\cal E} \otimes {\cal E} ')(\tilde \rho) \Bigr ), \end{eqnarray}
where the minimization is over all extensions $\tilde \rho$ of $\rho$ to
larger systems $ST$, and all possible quantum operations ${\cal E}'$ that
could occur on $T$. $F_1$ is a measure of how well the subsystem plus
its entanglement with the remainder of the system is stored. We minimize
over all possible extensions and dynamics for the remainder of
the system in order to obtain the {\em worst possible value} the fidelity
could have, regardless of the actual state or dynamics of the remainder of
the system. Clearly to understand error correction for small parts of a
quantum computer it would be desirable to 
use a quantity which depends only on the state of that part,
not on the state of the entire computer, and the quantity
$F_1$ is a natural candidate, since it measures the worst possible case.

A second quantity is also a useful measure of how well a system
plus entanglement is stored. It will turn out that this quantity is equal
to $F_1$. Define
\begin{eqnarray}
F_2(\rho,{\cal E}) := \min_{\tilde \rho} 
	F \Bigl ( \tilde \rho,
	({\cal E} \otimes {\cal I}_T)(\tilde \rho) \Bigr ). \end{eqnarray}
The motivation for this quantity is similar to that for $F_1$, except now
we assume that $T$ is subject to the identity dynamics ${\cal I}_T$, instead
of minimizing over all possible dynamics ${\cal E} '$ for $T$. The main use
of $F_2$ will be as an intermediate quantity.

We will now prove that $F_1$ and $F_2$ are equal. First, note that
\begin{eqnarray}
F_1(\rho, {\cal E}) \leq F_2(\rho,{\cal E}), \end{eqnarray}
since the minimization in $F_1$ clearly includes all the values being minimized
over for $F_2$. To see the reverse inequality, notice that
\begin{eqnarray}
F \Bigl ( \tilde \rho, ({\cal E} \otimes {\cal I}_T)(\tilde \rho) \Bigr ) \leq
	 F \Bigl ( ({\cal I}_S \otimes {\cal E} ')(\tilde \rho),
	 ({\cal E} \otimes {\cal E} ')(\tilde \rho) \Bigr ), \end{eqnarray}
by (\ref{non-decrease evop}), and thus
\begin{eqnarray}
F_2(\rho,{\cal E}) \leq F_1(\rho,{\cal E}). \end{eqnarray}
It follows that
\begin{eqnarray}
F_1(\rho,{\cal E}) = F_2(\rho,{\cal E}). \end{eqnarray}

Recently Schumacher \cite{Schumacher96a} introduced a quantity called
the {\em entanglement fidelity}, defined by the expression
\begin{eqnarray}
F_e(\rho,{\cal E}) := \bigl \langle \psi \big |
   ({\cal E} \otimes {\cal I})(|\psi\rangle \langle \psi|) \big | \psi
	\bigr \rangle, \end{eqnarray}
where $|\psi\rangle$ is any purification of $\rho$ (Schumacher proves that
any purification will give the same value), and $({\cal E} \otimes {\cal I})$
is the natural extension of the evolution operator to the
space on which $\rho$ has been purified.

We will show that the entanglement fidelity is equal to the
expressions $F_1$ and $F_2$ defined earlier.
In particular we prove that
\begin{eqnarray} \label{new fe}
F_e(\rho,{\cal E}) = F_2(\rho,{\cal E}) = \min_{\tilde \rho}
	F \Bigl ( \tilde \rho, ({\cal E} \otimes {\cal I}_T)(\tilde \rho)
	\Bigr ). \end{eqnarray}

The proof is as follows. Write $F_e := F_e(\rho,{\cal E})$ and
$F_2 := F_2(\rho,{\cal E})$.
The minimization in $F_2$ includes states $\tilde \rho =
|\psi\rangle \langle \psi|$ where $|\psi\rangle$ purifies $\rho$,
and thus
\begin{eqnarray}
F_2 & \leq & F \Bigl ( |\psi\rangle \langle \psi|,({\cal E} \otimes {\cal I})
   (|\psi\rangle \langle \psi|) \Bigr ) \nonumber \\
  & = & \bigl \langle \psi \big | ({\cal E} \otimes {\cal I})(|\psi\rangle
	 \langle \psi|) \big | \psi \bigr \rangle \nonumber \\
  & = & F_e. \end{eqnarray}
To show the reverse
inequality and thus complete the proof, suppose $\tilde \rho$ extends
$\rho$ to $ST$. Let $|\psi\rangle$ be a purification of $\tilde \rho$
on the space $STU$. It can be shown that
$({\cal E} \otimes {\cal I}_{TU})(|\psi\rangle\langle \psi|)$ is an
extension of $({\cal E} \otimes {\cal I}_T)(\tilde \rho)$ by using an
operator sum representation ${\cal E}(\rho) = \sum_i A_i \rho A_i^{\dagger}$,
as follows,
\begin{eqnarray}
\mbox{tr}_U & & \Bigl ( ({\cal E} \otimes {\cal I}_{TU})(|\psi\rangle \langle \psi|) \Bigr ) \nonumber \\
& & = \sum_i \mbox{tr}_U \Bigl ( (A_i \otimes I_T \otimes I_U)
	|\psi \rangle \langle \psi| (A_i^{\dagger} \otimes I_T \otimes I_U) 
	\Bigr ) \nonumber \\
	& & = \sum_i (A_i \otimes I_T) \tilde \rho (A_i^{\dagger} \otimes I_T)
	\nonumber \\
	& & = ({\cal E} \otimes {\cal I}_T)(\tilde \rho ).
\end{eqnarray}
Then from (\ref{non-decrease trace}) we see that
\begin{eqnarray}
F \Bigl ( \tilde \rho, ({\cal E} \otimes {\cal I}_{T})(\tilde \rho) \Bigr )
	 & \geq & F \Bigl ( |\psi\rangle \langle \psi|, ({\cal E} \otimes
	 {\cal I}_{TU}) (|\psi \rangle \langle \psi| ) \Bigr ) \nonumber \\
 & = & \bigl \langle \psi \big | ({\cal E} \otimes {\cal I}_{TU})(|\psi\rangle
	 \langle \psi|) \big | \psi \bigr \rangle. \end{eqnarray}
But since $|\psi\rangle$ is a purification of $\tilde \rho$ it is also
a purification of $\rho$ and thus
\begin{eqnarray}
F_e = \bigl \langle \psi \big | ({\cal E} \otimes {\cal I}_{TU})
	(|\psi\rangle \langle \psi|) \big | \psi \bigr \rangle. \end{eqnarray}
Combining the last two equations gives
\begin{eqnarray}
F \Bigl ( \tilde \rho,({\cal E} \otimes {\cal I} )(\tilde \rho) \Bigr ) \geq
	F_e. \end{eqnarray}
Minimizing the left hand side of this inequality over all extensions
$\tilde \rho$ of $\rho$ tells us that $F_2 \geq F_e$. Combining
this with the inequality $F_2 \leq F_e$ found earlier, and the
equality $F_1(\rho,{\cal E}) = F_2(\rho,{\cal E})$ gives the
final result,
\begin{eqnarray}
F_e(\rho,{\cal E}) = F_1(\rho,{\cal E}) = F_2(\rho,{\cal E}), \end{eqnarray}
as required.

The expression (\ref{new fe}) allows simple proofs of some of the
properties of the entanglement fidelity. For example, we see
immediately that
\begin{eqnarray}
F_e(\rho,{\cal E}) \leq F \bigl ( \rho,{\cal E}(\rho) \bigr ), \end{eqnarray}
since $\rho$ is a trivial extension of itself, and thus is included in the
minimization in (\ref{new fe}). This is an intuitively reasonable result;
it tells us that a state and its entanglement is not stored any better than
the state alone. In the earlier example concerning Bob and Alice,
the entanglement fidelity for Alice's system in the case of the first dynamics
is $F_e(\rho_A,{\cal E}_1) = 1$, whereas for the second dynamics it
is $F_e(\rho_A,{\cal E}_2) = \frac{1}{4}$, confirming our belief that
the first dynamics preserves the state plus entanglement well,
while the second dynamics preserves the state plus entanglement poorly.
It should also be noted that for pure states,
\begin{eqnarray}
F_e \Bigl ( |\psi\rangle\langle \psi |,{\cal E} \Bigr ) & = &
	\bigl \langle \psi \big | {\cal E} (
	|\psi\rangle \langle \psi|) \big | \psi \bigr \rangle \\
	& = &
	F \Bigl ( |\psi\rangle\langle \psi|, {\cal E}(
	 |\psi\rangle \langle \psi| )
	\Bigr ). \end{eqnarray}
That the fidelity and entanglement fidelity are equal for
pure states is intuitively reasonable, since pure states can not be entangled
with other systems, and thus no entanglement can be destroyed during the
storage process.

For completeness we will also mention the elegant explicit
formula for entanglement fidelity derived in \cite{Schumacher96a}. If
the quantum operation ${\cal E}$ is written in the form
\begin{eqnarray}
{\cal E}(\rho) = \sum_i A_i \rho A_i^{\dagger}, \end{eqnarray}
then it can be shown that
\begin{eqnarray}
F_e(\rho,{\cal E}) = \sum_i \mbox{tr}(A_i\rho) \mbox{tr}(A_i^{\dagger} \rho). \end{eqnarray}
This form allows the entanglement fidelity to be calculated explicitly
in actual examples.

A recent result due to Knill and Laflamme \cite{Knill96a} gives another
connection between the fidelity and entanglement fidelity. Theorem
5.3 of \cite{Knill96a} shows that if
\begin{eqnarray}
F \Bigl ( |\psi\rangle\langle \psi|,{\cal E}(|\psi\rangle\langle \psi|) 
	\Bigr ) \geq 1 - \epsilon \end{eqnarray}
for all pure states, $|\psi\rangle$, then
\begin{eqnarray}
F_e(\rho,{\cal E}) \geq 1 - \frac{3 \epsilon}{2}, \end{eqnarray}
for all states $\rho$. That is, if the fidelity is kept high
for all pure states, then it follows that the entanglement fidelity
is kept high for all states. 
This result shows that to preserve a quantum state and its entanglement
accurately, it is sufficient to keep the fidelity of storage high, provided
this is done for all pure states.

In this Rapid Communication two new expressions for the entanglement fidelity
have been obtained. These show that it is the entanglement fidelity
that is the important quantity to maximize in schemes for
quantum error correction. It may also be useful in applications such
as quantum teleportation, quantum cryptography and quantum coding in
which entanglement may need to be preserved.

\acknowledgments
I thank Howard Barnum, Carlton M. Caves, Christopher Fuchs
and Benjamin Schumacher for many enjoyable and enlightening discussions
about fidelity and quantum information. This work was supported in part
by the Phillips Laboratory (Grant No. F29601-95-0209) and
by a Fulbright Scholarship.

\end{document}